# Heterogeneities Shape Passive Intracellular Transport

P. Witzel[†], M. Götz[†], Y. Lanoiselée, T. Franosch, D. S. Grebenkov, and D. Heinrich[*]

[†] These authors contributed equally to this work.

[*] Corresponding author

Running Title: Heterogeneous Intracellular Transport

**ABSTRACT**

A living cell's interior is one of the most complex and intrinsically dynamic systems, providing an elaborate interplay between cytosolic crowding and ATP-driven motion, which controls cellular functionality. Here, we investigated two distinct fundamental features of the merely passive, not-bio-motor shuttled material transport within the cytoplasm of Dictyostelium discoideum cells: the anomalous non-linear scaling of the mean-squared displacement of a 150nm-diameter particle and non-Gaussian distribution of increments. Relying on single-particle tracking data of 320,000 data points, we performed a systematic analysis of four possible origins for non-Gaussian transport: (1) sample-based variability, (2) rare occurring strong motion events, (3) ergodicity breaking/ageing, and (4) spatio-temporal heterogeneities of the intracellular medium. After excluding the first three reasons, we investigated the remaining hypothesis of a heterogeneous cytoplasm as cause for non-Gaussian transport. A novel fit model with randomly distributed diffusivities implementing medium heterogeneities suits the experimental data. Strikingly, the non-Gaussian feature is independent of the cytoskeleton condition and lag time. This reveals that efficiency and consistency of passive intracellular transport and the related anomalous scaling of the mean-squared displacement are regulated by cytoskeleton components, while cytoplasmic heterogeneities are responsible for the generic, non-Gaussian distribution of increments.

**INTRODUCTION**

The dynamics of the cellular interior space comprise a sophisticated information and material transport regulating and controlling cellular functionality and thus enabling homeostasis. The cytoplasm of a living cell is a highly dynamic and complex system consisting of the densely crowded cytosol (1), the organelles and the structure-providing cytoskeleton components. Besides molecular diffusion on short ranges within the cytosol, vesicle transfer maintains macromolecular transport (2), immunological reaction, signal transduction (3,4), and provides nutrition. This transport consists of two distinct modes, as vesicles undergo alternating phases of directed transport via motor proteins and diffusion-like motion within the cytosol (5–7). In the first case, three types of motor proteins are involved (8): kinesins and dyneins travel along microtubules and allow for bidirectional transport towards the cell perimeter or the microtubule organizing center (MTOC) respectively (9,10), while myosins move along the actin cortex (11–13). This, so-called active or ATP-driven, vesicle transport enables long-distance cargo delivery within the cell. While our knowledge and understanding of this active vesicle transport advanced a lot within the last decades, the diffusion-like motion of vesicles still is not well understood. Though the relevance of passive vesicle transport in-between active bio-motor phases is unquestioned, especially due to its general necessity for finding binding sites, binding to other cellular compartments, and enhancing of search strategies (14), the experimental data basis in literature for these processes is scarce. Thus, we investigate this particular transport phenomenon via single-particle tracking of fluorescent 150 nm-diameter nanoparticles that mimic the dynamics of vesicles. With this diameter we match the intermediate range of vesicles in most cell types and exceed the average actin mesh size of 100 nm (15), so particles do not enter the dense mesh located at the cell membrane and do not get stuck within. The particles exhibit no functional surface coating to prevent molecular motor binding and to ensure purely passive transport.

We utilized the cell system *Dictyostelium discoideum* (*D. discoideum*), a social amoebae, due to their similarities to mammalian cells regarding cell behavior, structure and intracellular signaling, with a completely sequenced genome and a simple and basic cytoskeleton structure (16–19). *D. discoideum* serves as a relevant model organism for amoeboid migration (20,21), chemotaxis (22,23), phagocytosis, and human diseases (17,24). Of special interest for the investigations on intracellular processes is the truly three dimensional shape of *D. discoideum* cells, which represents the natural morphology of cells in physiological 3D environments in the human body, in contrast to adherent, flat mammalian cells in conventional *in vitro* cell culture on flat surfaces (25). The cytoskeleton of *D. discoideum* cells is simple but comparable to mammalian cells and consists mainly of the actin cortex and microtubules, two biopolymers determining cell shape (see upper left part of Fig. 1a). While the actin cortex is a reorganizing network that is the densest at the plasma membrane, the dynamically sweeping microtubules originate from the microtubule organizing center (MTOC) adjacent to the nucleus, spreading towards the membrane and functioning as rails for intracellular transport by motor proteins (26,27). A relevant advantage of *D. discoideum* is the possibility to depolymerize these cytoskeleton components almost completely by drug treatment, without causing cell death (see Supplemental Information). Here, the drug latrunculin A prevents formation of the actin cortex by binding to the G-Actin monomers (28); analogue, benomyl and nocodazole impede polymerization of the microtubule monomers tubulin (29,30). The injection of nanoparticles directly into the cytosol avoids their encapsulation by internal vesicles. As these particles are functionally uncoated, they cannot bind to molecular motors (31) and thus witness uniquely the passive intracellular transport.

With this setup, we investigate two distinct fundamental aspects of the passive (not actively bio-motor shuttled) vesicle-like transport within the cytoplasm: the anomalous non-linear scaling of the mean-squared displacement (MSD) and the non-Gaussian distribution of its increments. While the anomalous scaling and related subdiffusive and superdiffusive motions of various tracers in living cells have been thoroughly studied in the last twenty years (32–39), the non-Gaussian feature has only recently been experimentally discovered for subdiffusive motion of RNA-protein particles in live *Escherichia coli* and *Saccharomyces cerevisiae* cells (40). The observed exponential (Laplace) distribution was attributed to a combination of an exponential distribution in the time-averaged diffusivities and non-Gaussian behavior of individual trajectories due to significant heterogeneity between trajectories and dynamic heterogeneity along single trajectories. Similar results were obtained for the acetylcholine receptors diffusing on live muscle cell membranes (41). In turn, such features remain unknown for passive vesicle transport inside living cells due to the difficulty of acquiring enough sufficiently long trajectories of tracers in the living cell interior. Relying on a large amount of single-particle tracking data with in total 320,000 data points, we uncover the origin of non-Gaussian intracellular motion based on the analysis of increment probability densities, the ergodicity breaking parameter, and autocorrelation functions. We show that the cytoskeleton components, the actin cortex and the microtubules, control the efficiency of passive intracellular transport but not the non-Gaussian increment distributions. Unexpectedly, we find that a rescaling of the increment distributions by the mean increment results in a collapse of all increment distributions onto one single master curve with an exponential tail. This master curve is independent of the cytoskeleton state and the lag time. We are able to assign the cause of non-Gaussian transport to spatio-temporal heterogeneities within the cytoplasm, opening new ways for future modeling of passive intracellular vesicle dynamics.

**MATERIALS AND METHODS**

**A. Materials**

$Na_2HPO_4 \cdot 2H_2O$ (≥ 99,5 % p.a., Carl Roth GmbH + Co. KG, Germany), $KH_2PO_4$ (≥ 99 % p.a., Carl Roth GmbH + Co. KG, Germany), D(+)-Maltose Monohydrate (≥ 95 % p.a., Carl Roth GmbH + Co. KG, Germany), Latrunculin A (ThermoFischer Scientific, Germany), Methyl 1-(butylcarbamoyl)-2-benzimidazolecarbamate (Benomyl, 95 %, Sigma-Aldrich, Germany), Methyl N-(5-thenoyl-2-benzimidazolyl)carbamate (Nocodazole, ≥ 99 %, Sigma-Aldrich, Germany), Gentamycin (G-418, Biochrom AG, Germany), Blasticidin (Blasticidin S hydrochloride, Sigma Aldrich Chemie GmbH, Germany), the cell culture medium HL5-C (ForMediumTM, United Kingdom) and nano-screenMAG-D particles (150 nm, poly dispersion index PDI 0.14-0.17, ChemiCell, Germany) were used as received without further purification.

**B. Cell Culture**

*Dictyostelium discoideum* cells from AX2 strain as wild type (WT), with LimEΔcc-GFP expressed in LimEΔcc-null and with alpha tubulin-GFP expressed in LimEΔcc-mRFP were provided by Dr. Günther Gerisch (Max-Planck Institute of Biochemistry, Germany). Cells were grown at 21 °C in the cell culture medium HL5 adjusted to pH = 6.7. For the LimEΔcc-GFP expressed in LimEΔcc-null and alpha tubulin-GFP expressed in LimEΔcc-mRFP strains the medium was complemented by the antibiotics Gentamycin and Blasticidin at a concentration of 10 µg mL$^{-1}$, respectively. The cells were kept in the vegetative state at a confluence under 40 %.

**C. Experimental Procedure**

To avoid fluorescence of the medium, the cells were transferred to a non-fluorescent medium of 3.57 mM Na$_2$HPO$_4$·2 H$_2$O, 3.46 mM KH$_2$PO$_4$ and 52.58 mM D(+)-maltose monohydrate in water, adjusted to pH = 6.7, 2 h before the experiment was started. Subsequently, green fluorescent nano-screenMAG-D particles (a priori agitated by sonication for 10 min) with a diameter of 150 nm were added to the cell suspension at a final concentration of 0.36 mg mL$^{-1}$. These particles exhibit a magnetite core surrounded by a starch matrix and offer no binding sites to molecular motors like kinesins, dyneins, or myosins. To prevent transport of the particles in vesicles within the cytosol, ballistic injection was performed by centrifugation at 4,000 rpm in four subsequent periods of 3, 4, 4 and 5 min duration, each followed by gentle agitation for 5 min. For imaging, the cells were settled in an 8-well plate with glassy bottom (ibidi, Germany). After a resting time of 30 min, allowing the cells to adhere to the ground, excess nanoparticles were removed by washing the cells for three times with non-fluorescent medium. The inhibition of the actin polymerization (noAct) was performed by adding latrunculin A to the cell suspension at a final concentration of 20 µM 30 min before the start of measurements (31). To prevent microtubule polymerization (noMT), the agents benomyl at a concentration of 50 µM and nocodazole at a concentration of 20 µM were mixed and added to the cells 30 min before the beginning of measurements (31). For the case without both microtubules and actin cortex (noCyt), latrunculin A, nocodazole and benomyl were added to the cells 30 min prior to the experiments at the above mentioned concentrations. To monitor the drug efficiency, *D. discoideum* cells with LimEΔcc-mRFP expressed in alpha tubulin-GFP have been treated equally at the same time. The fluorescent signals of GFP for the microtubules and RFP for Lim (a protein present at sites of freshly polymerized actin) vanished by treatment with benomyl/nocodazole and latrunculin A, or both, respectively. After removing the drugs by washing the cells with medium, the cells regenerated fully, showed normal behavior and cell division, with the fluorescent signals for both the Lim protein, indicating actin polymerization, and the microtubules visible again.

**D. Microscopy**

Live cell imaging was performed on a Nikon eclipse Ti-E microscope (Nikon, Japan) equipped with an EM-CCD camera (C9100-50, Hamamatsu, Germany) using a 100x oil immersion objective with a NA of 1.4 (Nikon, Japan) and an additional inbuilt lens of 1.5x magnification yielding a resolution of 54 nm per pixel. An Intensilight (Nikon, Germany) and a filter set (F36 525 HC, AHF, Germany) with transmission wavelengths from 457 nm to 487 nm for excitation and transmission wavelengths from 500 nm to 540 nm for the detection path were utilized to image the fluorescent nanoparticles. These nanoparticles within *D. discoideum* cells were imaged every 49 ms adjusting the focal plane manually with the acquisition software NIS-Elements AR at a 14 bit image depth. Imaging was performed at intermediate cell height and the acquisition took place always close to the central cell plane and at similar distances from the nucleus and the outer cell periphery, to keep locus-specific differences of the nanoparticle motion as small as possible. Particle tracking was performed by tracking software, OpenBox 1.64 (Informationssysteme Schilling, Germany), using 2D Gaussian fits. This setup allows for a lateral tracking accuracy of ± 10 nm (42).

Videos S1 and S2 for the visualization of the cytoskeletal dynamics of LimEΔcc-GFP expressed in LimEΔcc-null and alpha tubulin-GFP expressed in LimEΔcc-mRFP strains were acquired on a confocal microscope setup equipped with a spinning disk unit. The setup consists of an inverted microscope (Ti-E, Nikon, Japan), a spinning disk unit (UltraVIEW™ VoX, PerkinElmer, USA), a CCD camera (C9100-50 Hamamatsu, Germany) and two lasers (488nm and 561 nm, Yokogawa Electric Corporation, Japan). The imaging was performed by a 100x immersion oil objective with a NA of 1.45 at a frame rate of 0.5 fps for Video S1 and 10 fps for Video S2. For acquisition the software Volocity (PerkinElmer, USA) was used.

**E. Statistical Analysis**

All statistical analyses are applied to the unchanged two-dimensional single-particle trajectories, i.e. the successive positions $\{(x_m(n\tau), y_m(n\tau))\}$ $(n = 1, \ldots, N_m)$ of the *m*-th tracer acquired at equal time intervals of duration $\tau$ (here $N_m$ is the number of points of the *m*-th trajectory). We are well aware of localization errors as well as the spurious correlation in the velocity autocorrelation (VAF) appearing due to various measurement noises. While advanced estimators can correct for some errors in case of Brownian motion (43–45), no generic correction protocol is available for a more general, anomalous, non-Gaussian motion, as in this study. Moreover, as our goal is not the estimation of the scaling exponent but the analysis of the increments distributions, we keep using the genuine data sets to avoid eventual biases or artifacts.

The empirical probability density $P(r, t)$ of absolute magnitudes $r$ of one-dimensional projected increments was evaluated at discrete points $r = k\rho$ (with a chosen bin size $\rho$) and $t = n\tau$ as $P(k\rho, n\tau) = \frac{p_k(n\tau)}{\rho\, p(n\tau)}$, where $p_k(t)$ is the number of times that the absolute one-dimensional increment $|x_m(q\tau + n\tau) - x_m(q\tau)|$ or $|y_m(q\tau + n\tau) - y_m(q\tau)|$ belongs to the interval $(k\rho, (k+1)\rho)$ and the normalization factor $p(n\tau)$ is the sum of all $p_k(n\tau)$. Here, one uses the increments for all tracers in a cytoskeleton state (i.e. $m = 1, \ldots, M$) and for all possible shifts along each trajectory (i.e. $q = 1, \ldots, N_m - n$). In other words, the probability density of increments includes the ensemble average over all trajectories in a state (e.g. the WT case) and the time average along each trajectory. The time average was crucial to improve the statistics of increments. For the same purpose, the increments along *X* and *Y* coordinates were merged. From all trajectories in each of four cytoskeleton states (Table S1), we had 152,804, 160,958, 176,528, and 164,300 one-step increments for the WT, noAct, noMT, and noCyt cases, respectively. This large amount of data points is necessary for a sufficient signal-to-noise ratio of the increment probability density distributions and to average locus-specific variations of the particle motion. Based on these increments, histograms were produced for each cytoskeleton state and at different lag times $t$, from $\tau$ to $100\,\tau$.

In addition, the time-averaged velocity autocorrelation function (VAF), characterizing correlations of one-step one-dimensional increments, was computed for each trajectory ($m = 1, \ldots, M$) as

$$C_m(n\tau) = \frac{1}{2}\left(\frac{U_m^x(n)}{U_m^x(0)} + \frac{U_m^y(n)}{U_m^y(0)}\right) \quad (3)$$

where

$$U_m^x(n) = \frac{1}{N_m - n}\sum_{k=1}^{N_m - n}\left(x_m(k\tau) - x_m((k-1)\tau)\right)\left(x_m((k+n)\tau) - x_m((k+n-1)\tau)\right)$$

is the time-averaged VAF of one-step increments along $X$ coordinate (similarly $U_m^y(n)$ for the $Y$ coordinate), $N_m$ is the number of points of the m-th trajectory, and $n$ is the discrete lag time. Note also that

$$\sigma_m = \left(\frac{U_m^x(0) + U_m^y(0)}{2}\right)^{\frac{1}{2}} \quad (4)$$

is the empirical standard deviation of one-step one-dimensional increments (Supplemental Table S2) that was used to obtain the rescaled probability densities.

The empirical histograms were fitted by a standard non-linear Levenberg-Marquardt least-squares algorithm in Matlab (lsqcurvefit, The MathWorks, Natick, MA). Fitting was performed on a trustworthy region of the increment probability density distributions, defined by equal or more than 100 data points in the interval $(k\rho, (k+1)\rho)$.

The distributed model to fit the scale function $P(r,t)$ is derived in the Supplemental Information.

**RESULTS AND DISCUSSION**

Passive intracellular material transport was probed by single-particle tracked tracers of a diameter of 150 nm. These nanoparticles (one per cell) have been injected directly into the cytosol of *D. discoideum* cells by centrifugation – not enclosed by vesicles – and thus reflect the passive dynamics of the cytoplasmic space exclusively. To elucidate the cytoplasm influence, *D. discoideum* cells were drug treated for four different cytoskeleton states (Fig. 1a): untreated wild type (WT), depolymerized microtubules (noMT), depolymerized actin cortex (noAct), and with disrupted cytoskeleton (noCyt) by combination of polymerization inhibiting drugs (see Supplemental Information). For all experiments, cell-mediated, bio-motor regulated transport, e.g. along microtubules, has been effectively excluded by using particles that are unable to bind to cellular components. Based on more than 320,000 data points of nanoparticle trajectories (insets of Fig. 1b–e, Supplemental Fig. S1), we analyzed the probability densities of magnitudes of one-dimensional increments for different lag times to gain a deeper insight into the cause for non-Gaussian intracellular transport.

## A. Probability densities of nanoparticle displacement

For all four distinct cytoskeleton states, these probability densities (Fig. 1b–e) exhibit an overall non-Gaussian behavior comprising a parabolic Gaussian-like region for small increments and an exponential tail for larger increments. This behavior is different from the exponential (Laplace) distribution without parabolic region, observed for the subdiffusive motion of RNA-protein particles (40). One difference between Gaussian and exponential distributions is that large increments are more likely to be realized in the latter case. In both cases of deprived actin (noAct, noCyt), further deviations from the exponential tail at even larger increments can be noticed. Here, the dense actin network (see Supplemental Video 1) is absent leading to a broader increment distribution as compared to the WT state (Fig. 1c) due to less restricted particle motion (Fig. 1b). In contrast, single-particle tracking in cells lacking microtubules (noMT) yields less spread trajectories (inset of Fig. 1d), where the corresponding probability densities (Fig. 1d) behave similarly to those of the WT state yet at smaller increments. In the noMT state, the cell is set to a state of pure crowding due to the lack of the cytoplasm-fluidifying effect arising from the microtubules' sweeping motion (see Supplemental Video 2). Depolymerizing the actin cortex and microtubules in the no-cytoskeleton state (noCyt) combines these effects, obvious at longer lag times (Fig. 1e): the cells lack both the confinement by the actin cortex and the active stirring of the microtubule motion. This results in a constrained particle motion as compared to the WT state, but shows larger increments than in the noMT state, where the additional confining effect of the actin cortex is present. This increment probability density analysis reveals the opposing functions of both cytoskeleton components in the cytoplasmic space of a living cell: while the actin cortex confines material transport within the cell interior, the actively driven dynamics of the microtubules fluidifies the cytoplasm, fighting against intracellular crowding (see also the MSD in Fig. 3c and discussion below).

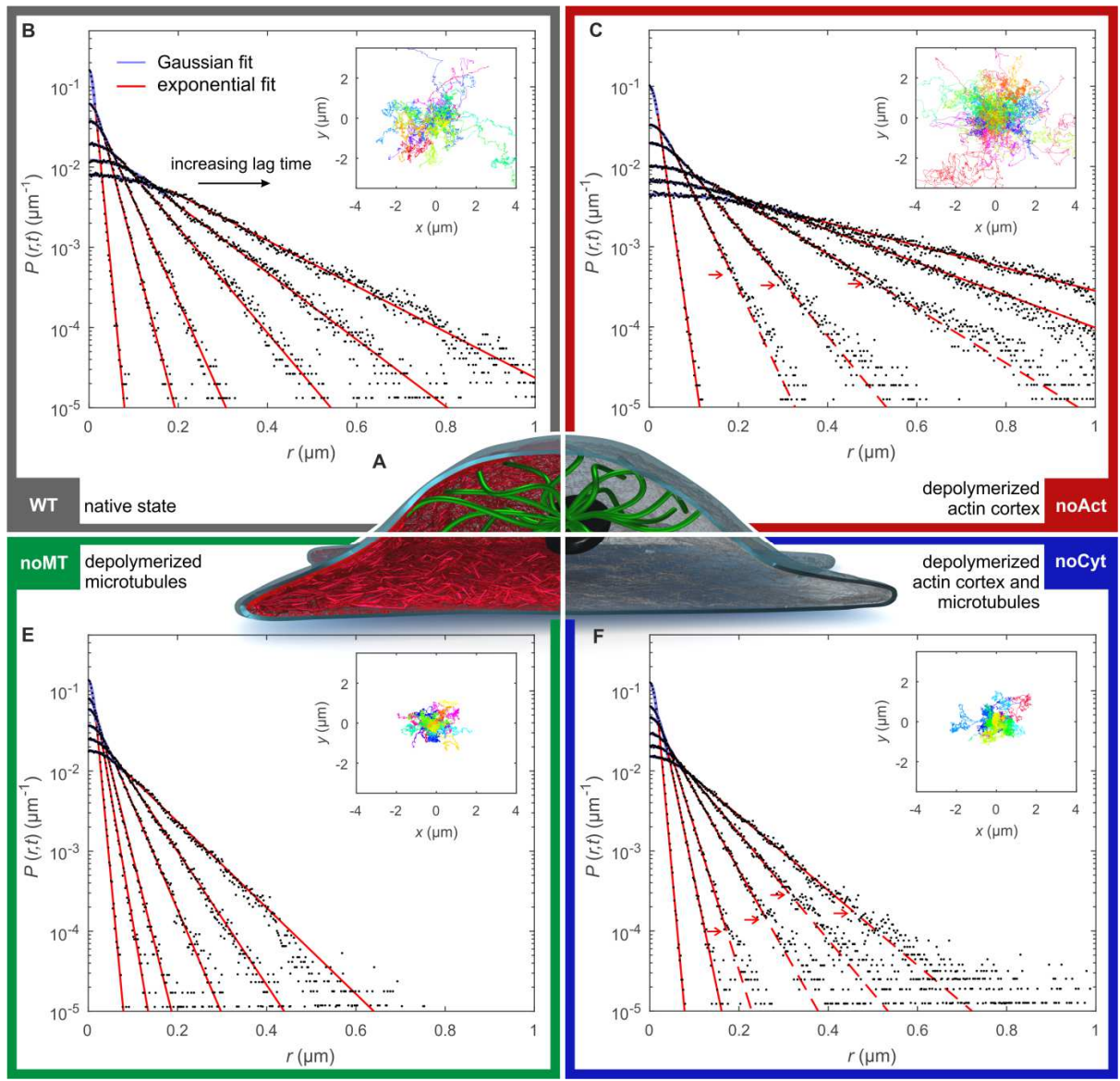

**Figure 1 | Increment probability densities of intracellular nano-tracer trajectories in four different cytoskeleton states.** (**a**) Illustration of a living eukaryotic cell exhibiting depolymerized cytoskeleton components (grey: membrane, black: nucleus, green: microtubule organizing center with microtubules, red: actin cortex). (**b-e**) Probability densities of absolute magnitudes of one-dimensional projected increments, $P(r,t)$, of intracellular nano-tracers for lag times $t = \tau, 5\tau, 10\tau, 25\tau, 50\tau, 100\tau$ with the inverse frame rate $\tau = 49$ ms for four different cytoskeleton states: (**b**) wild type (WT), (**c**) depolymerized actin cortex (noAct), (**d**) depolymerized microtubules (noMT) and (**e**) depolymerization of both the actin cortex and the microtubules (noCyt); solid lines show the fit of a Gaussian function (light blue) followed by an exponential fit (red). For large lag times in the actin deprived cases (noAct and noCyt) a deviation from the exponential fit (red) is indicated by dashed lines, arrows show the transition point; (insets) all acquired trajectories of nano-tracer motion displayed in different colors, all started from the origin (see details on the trajectories in the Supplemental Information).

## B. Cause of non-Gaussian dynamics in passive intracellular transport

Material transport behavior with non-Gaussian increments has been observed and investigated in several non-living, complex systems (46–62) ranging from glasses to granular matter and artificial F-actin networks, as well as for the dynamics of receptors on live cell membrane (41) and recently in living cell interiors (63,40). In those systems, non-Gaussian transport features have been rationalized either by (1) sample-based variability, (2) rare occurring strong motion events, (3) ageing or (4) spatio-temporal heterogeneities of the medium. In the following, we investigate which of these reasons is applicable to the eukaryotic intracellular medium.

*1. Sample-based variability*

Non-Gaussian increment distributions across an ensemble of trajectories can be interpreted in two ways: (i) each trajectory exhibits normally distributed increments, whereas their diffusivity changes from one trajectory to another, or (ii) the diffusivity changes along each trajectory. The validation of the Gaussian hypothesis for increments from each trajectory is restricted due to a limited length of individual trajectories, prohibiting an accurate computation of the probability density of increments for each single trajectory. For this reason, we resort to statistical tests of Gaussianity, which can be applied to individual trajectories. Among various statistical tests, we chose the Anderson-Darling and the Shapiro-Wilk tests, both shown to be the most efficient for testing Gaussianity on small samples (64,65). This analysis (see Supplemental Table S2) reveals that non-Gaussian features are intrinsic to each trajectory. To reduce the general cell-related variability, all one-step increments of all trajectories were rescaled by dividing the one-step increments within each individual cell by their standard deviation. Fig. 2a–c shows the rescaled distributions for three lag times $t$: $\tau$, 10 $\tau$, and 100 $\tau$, where $\tau$ = 49 ms is the one-step duration, which represents the inverse frame rate. For all cytoskeleton states, the rescaled probability densities collapse to a single curve at the smallest lag time, whereas the curves are paired at a longer lag time of t = 100 $\tau$: the WT and noAct states exhibit larger increments caused by the presence of the microtubules' sweeping motion, whereas the noMT and noCyt states show smaller increments arising from intracellular motion without the influence of microtubule dynamics. Furthermore, this rescaling eliminates deviations from the exponential tail at large increments (Fig. 2a–c), which occurred exclusively in the noAct and noCyt states, both characterized by a depolymerized actin cortex (Fig. 1c,e). Thus, the original deviations can be attributed to sample-based variability among individual cells, also evident from the larger spread of the one-step increment standard deviations between individual cells in both actin deprived cases (noAct and noCyt) (see Supplemental Tables S3 and S4). This elucidates the important role of the actin cortex as reducing the cell-to-cell heterogeneity of the cellular interior space for well-regulated, consistent transport dynamics across a cell population.

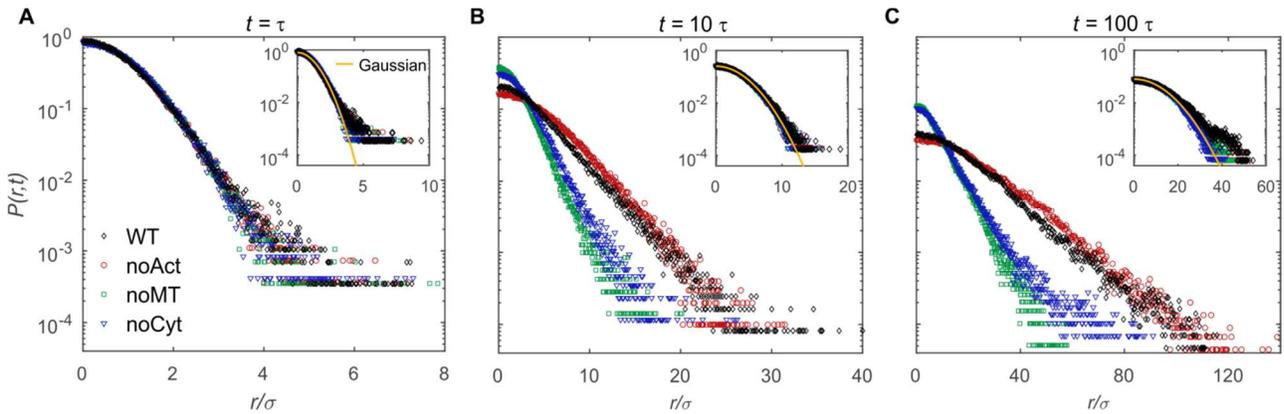

**Figure 2 | Increment probability densities of intracellular nano-tracer trajectories rescaled by standard deviation.** Probability densities of increments for four cytoskeleton states (WT, noAct, noMT, noCyt) for lag times $t = \tau$ (a), 10$\tau$ (b) and 100$\tau$ (c) with the inverse frame rate $\tau$ = 49 ms, the one-step increments were rescaled by the standard deviation $\sigma$ for each individual cell. Insets show the increment probability densities for randomly shuffled, rescaled increments for lag times $t = \tau$ (inset a), 10$\tau$ (inset b) and 100$\tau$ (inset c) with the inverse frame rate $\tau$ = 49 ms, solid line shows the one-sided Gaussian distribution $2\exp\left(\frac{-r^2}{2\sigma^2 t}\right)/\sqrt{2\pi\sigma^2 t}$, with $\sigma$ = 1 due to the rescaled increments. (Fig. 2a and its inset are identical, because shuffling of one-step increments does not affect their distribution.)

At the same time, the persistence of an exponential tail in the rescaled data of all cytoskeleton states eliminates sample-based variability as a reason for non-Gaussian features in passive intracellular motion.

*2. Rare occurring strong motion events*

Another cause for non-Gaussian increment distributions could be strong, rare events. In the cytosol, these can arise from nearby passing microtubules, intracellular cargo, or actin reorganization. For the cell state with both depolymerized microtubules and depolymerized actin cortex (noCyt), all active cytoskeleton dynamics within the cell have been eliminated, but the increment distribution still exhibits non-Gaussian behavior (Fig. 2a–c, blue triangles).The existence of a non-Gaussian increment distribution with an exponential tail in the noCyt case – without any cytoskeleton components – clearly evidences against the second option of rare occurring strong events.

*3. Ageing*

Possible ageing effects, i.e. non-vanishing dependence on initial conditions and progressive slowing down of the dynamics, can also lead to non-Gaussian transport. In fact, long stalling periods of a particle may render its increments at a fixed lag time highly heterogeneous and strongly correlated, breaking the applicability of the central limit theorem. These effects have been theoretically studied within a continuous-time random walk model (66), whereas the ageing and ergodicity breaking of the receptor dynamics on a cell membrane were recently reported (67,68). Ageing effects can be characterized by the ergodicity breaking parameter (69,70) that describes fluctuations of the time-averaged mean-squared displacement. As Figure 3a shows, the ergodicity breaking parameter vanishes with increasing experimental observation time, which corresponds to the length of trajectories, so that an ageing effect can be excluded. Thus, the motion of 150 nm particles within the cytosol is ergodic at investigated time scales and ageing cannot account for non-Gaussian increment distributions (see Supplemental Information).

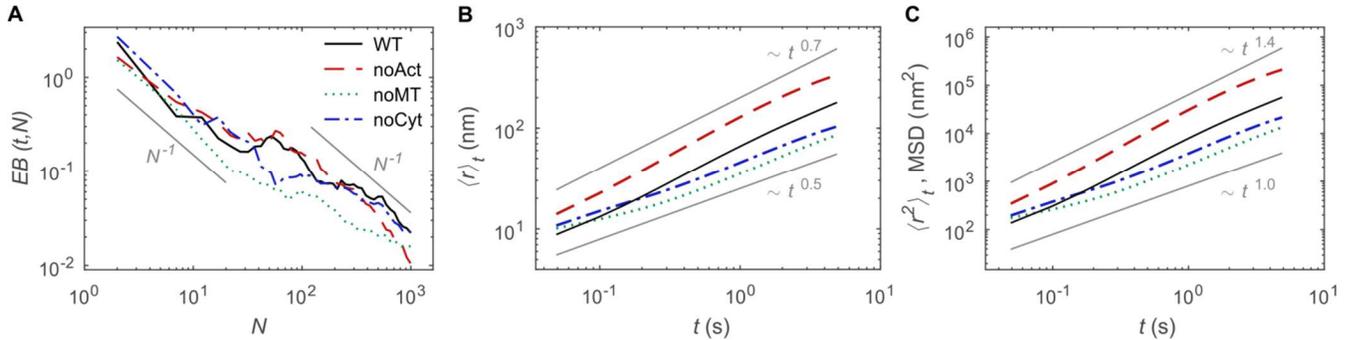

**Figure 3 | Ergodicity breaking parameter, mean absolute increment and mean-squared displacement** for four cytoskeleton states (WT, noAct, noMT, noCyt) (**a**) Ergodicity breaking parameter (EB) as function of trajectory length N, two indicative lines in grey show the typical 1/N decay for Brownian motion. (**b**) Mean absolute increments $\langle r \rangle_t$ as functions of lag time $t$ for all cytoskeleton states, grey lines show two power laws: $t^{0.5}$ for diffusive and $t^{0.7}$ for super-diffusive motion. (**c**) Mean-squared displacement (MSD) as functions of lag time $t$ for all cytoskeleton states, grey lines show two power laws: $t^{1.0}$ for diffusive and $t^{1.4}$ for super-diffusive motion.

## 4. Spatio-temporal heterogeneities of the medium

After excluding three possible reasons of non-Gaussian behavior within the intracellular medium, we now address the hypothesis of a spatio-temporal heterogeneous medium. Non-Gaussian distributions of increments are found for all considered lag times and all cytoskeleton states (Fig. 1 and 2). For lag time $t = 1\ \tau$, a possible reason for non-Gaussianity could be that one-step increments are not identically distributed. However, the distribution of large lag time increments, which could be considered as the sum of numerous one-step increments, would be unavoidably Gaussian due to the central limit theorem (i.e. according to Lindeberg condition), if the one-step increments were independent. To prove this non-independence, we perform a random re-shuffling of one-step increments within each experimental trajectory to destroy all possible correlations (insets of Fig. 2). In all four cytoskeleton states, the re-shuffled increment distributions become much closer to each other and to a Gaussian curve with moderate or minor deviations. The non-Gaussian character of the original (non-re-shuffled) increment distributions at large lag times could thus be rationalized by the presence of strong, long-range correlations between one-step increments. To verify this hypothesis, the velocity autocorrelation function (VAF) is evaluated, which reflects the temporal change in the magnitude and directionality of successive increments. The VAF reveals positive correlations over time scales of about 2 seconds only for the WT and the noAct cases (Fig. 4a), which can thus be attributed to sweeping motion of microtubules (see Supplemental Video 2). These correlations could be a possible reason for non-Gaussian increment distributions at large lag times, but only for these two cell states (WT and noAct). In the cases without microtubules (noMT and noCyt), only short-ranged, negative correlations on a sub 100 ms time scale are present and thus not sufficient to rationalize the observed non-Gaussian increment distributions in these cell states.

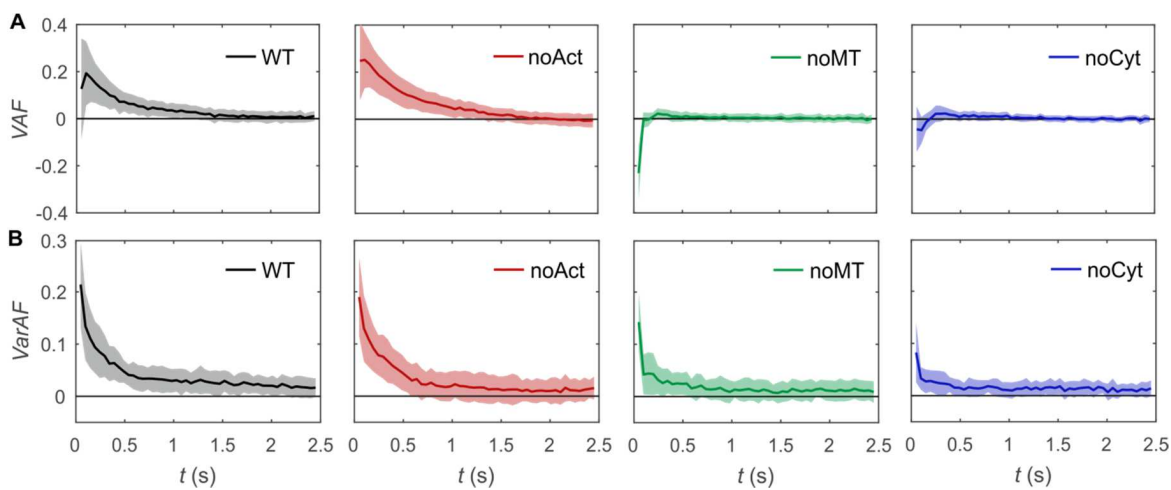

**Figure 4 | Velocity and variance autocorrelation functions** (a) Time-averaged velocity autocorrelation functions (VAF) for four cytoskeleton states (WT, noAct, noMT, noCyt), solid lines correspond to the mean value obtained by averaging the velocity autocorrelations of all trajectories, shaded areas mark the standard deviation.(b) Time-averaged variance autocorrelation functions of squared increments (VarAF) for four cytoskeleton states (WT, noAct, noMT, noCyt), solid lines correspond to the mean value obtained by averaging the variance autocorrelations of all trajectories, shaded areas mark the standard deviation.

If the dynamics was Gaussian, the VAF would fully determine its correlation properties. In turn, for a non-Gaussian process, the VAF essentially reflects correlations in the directions of increments but is only weakly sensitive to eventual correlations in the magnitudes of increments. We therefore analyze such higher-order correlations by introducing the variance autocorrelation function (VarAF) of squared increments (see Supplemental Information and (71)). Figure 4b shows positive, slowly decreasing autocorrelation of squared increments for all cytoskeleton states. A plausible scenario for obtaining correlated squared increments is a change of the effective diffusivity along the nanoparticle trajectory, arising either by a slowly evolving environment around the tracer (72,73,71,74–76) or by the exploration of a quasi-static but spatially heterogeneous cytoplasm (77,78), or both. These two distinct scenarios correspond to random motion with either time- or space-dependent diffusion coefficients, or again both. A direct experimental proof for heterogeneities causing non-Gaussian increment distributions within the cellular interior is too challenging with the current technological state-of-the-art and the need for a large amount of data points. Our indirect experimental evidence of medium-induced spatio-temporal correlations, influencing intracellular transport, illustrates that the analysis of increment distributions provides deeper and more versatile insights onto the intracellular dynamics in comparison to former studies based solely on mean-squared displacement or velocity autocorrelations. When the cytoskeleton elements are progressively removed, the local diffusivity autocorrelations are getting lower, from the WT and noAct states to noMT and noCyt states. This qualitative observation agrees with the expected reduction of heterogeneities in these cytoskeleton states.

## C. Generic character of passive intracellular transport

An additional rescaling of the *n*-step increments – already rescaled by the individual standard deviation for every cell – was performed by the mean increment, $\langle r \rangle_t$, over all trajectories within a cytoskeleton state at lag time $t = n\tau$. This resulted in a superimposition of the probability densities for all cytoskeleton states and lag times onto one master curve (Fig. 5 and see Supplemental Fig. S3 in for each cytoskeleton state). Thus, these probability densities of magnitudes of increments exhibit a scaling form,

$$P(r,t) = \langle r \rangle_t^{-1} \hat{P}(\hat{r}), \qquad \hat{r} = \frac{r}{\langle r \rangle_t} \qquad (1)$$

independent of the lag time *t* and independent of whether the motion is influenced by active cytoskeleton components – microtubules or actin cortex – or not. Note that the inverse increment $\langle r \rangle_t^{-1}$ in equation (1) accounts for the dimensionality of the probability density. Residual deviations from the master curve appear at large increments when the probability density is small and thus characterizes relatively rare moves. These deviations are stronger for the noCyt case at t = 100 τ that might be related to larger heterogeneity due to the lack of actin.

The scaling form (1) clearly indicates a generic intracellular dynamics feature, which is independent of the cytoskeleton components in *D. discoideum* cells for the used 150 nm-diameter particles and for lag times up to 4.9 s. Based on this scaling behavior, the transition of the increment probability distributions from a Gaussian-like regime to an exponential tail is an intrinsic property of the underlying intracellular material transport, and the shape of the probability density is captured solely by the master curve $\hat{P}$. Thus, microtubules and the actin cortex do not change the increment distribution, which is controlled by heterogeneities, but regulate the efficiency of the intracellular dynamics like a gear shift mechanism, enabling well-controlled intracellular material and information transport.

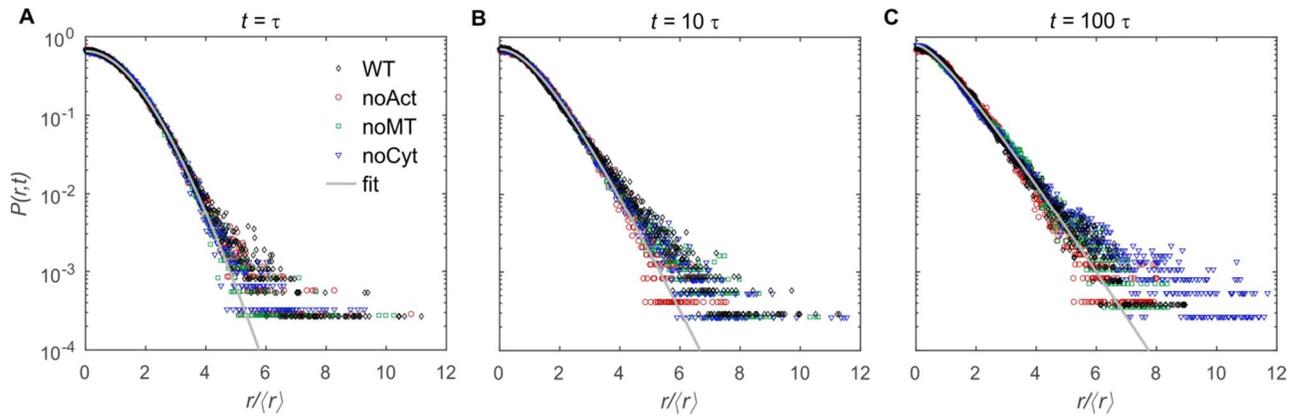

**Figure 5 | Increment probability densities of intracellular nano-tracer trajectories rescaled by mean increment** (a) Probability densities of increments for four cytoskeleton states (WT, noAct, noMT, noCyt) for lag times $t = \tau$ (**a**), $10\tau$ (**b**) and $100\tau$ (**c**) with the inverse frame rate $\tau$ = 49 ms with additional rescaling of the *n*-step increments by their mean absolute increments $\langle r \rangle_t$ at lag time $t = n\tau$ for each cytoskeleton state; solid lines show a representative fit by equation (1) and (2) for the noMT case, fits for the three other cases are overlapping and therefore not shown. The values of the shape parameter ν are 3.5 (a), 3.4 (b), and 1.8 (c), see also Supplemental Figure S2 for the dependence of ν on the lag time for all four cytoskeleton states.

### D. Two-parameter model

To gain insight into the behavior of the master curve $\hat{P}$, we propose a distributed model of Gaussian increments with random standard deviations σ exhibiting only two parameters. An explicit form of the master curve $\hat{P}$ is deduced (see Supplemental Information for more details) from the assumed chi distribution for $\sigma$:

$$\widehat{P}(\hat{r}) = \frac{\alpha(\alpha\hat{r})^{\nu-\frac{1}{2}} K_{\nu-\frac{1}{2}}(\alpha\hat{r})}{\sqrt{\pi}\, 2^{\nu-\frac{3}{2}} \Gamma(\nu)}, \quad \alpha = \frac{2\,\Gamma\left(\nu+\frac{1}{2}\right)}{\sqrt{\pi}\,\Gamma(\nu)}, \tag{2}$$

where $K_{\nu-\frac{1}{2}}$ is the modified Bessel function of the second kind, $\Gamma(\nu)$ denotes the Gamma function, and the shape parameter $\nu$ quantifies the spread of standard deviations $\sigma$. At the particular value $\nu = 1$, our model yields the one-sided exponential (Laplace) distribution, $\widehat{P}(\hat{r}) = \frac{1}{2}\exp(-\hat{r})$, observed in other experiments (40). The shape parameter $\nu > 1$ is needed to reproduce a parabolic region at small increments observed in our experiments. The second parameter is the scale of increments, $\langle r \rangle_t$, which incorporates the scaling with the lag time. The resulting probability density $P(r,t)$ reproduces both the Gaussian-like regime at small increments and the exponential tail at large increments, in perfect agreement with experimental data for all cytoskeleton states and all lag times (Fig. 5).

The collapse of all increment probability densities onto a single master curve implies that the dynamics-related differences between the four cytoskeleton states are captured solely by the mean absolute increment $\langle r \rangle_t$. Figure 3b shows $\langle r \rangle_t$, as well as the closely related MSD $\langle r^2 \rangle_t$ (Fig. 3c), as a function of the lag time $t$. In cases without microtubules (noMT and noCyt cases), one observes a linear scaling of the MSD, $\langle r^2 \rangle_t \propto t$ for larger lag times, while initial subdiffusive regimes are present with $\langle r^2 \rangle_t \propto t^{0.6}$ up to 0.1 seconds for the cytoskeleton state without microtubules (noMT) and with $\langle r^2 \rangle_t \propto t^{0.9}$ up to a lag time of 0.3 seconds in cells with depolymerized cytoskeleton (noCyt). We emphasize that super- or subdiffusive scaling of the MSD with the lag time and the non-Gaussian character of the increment distribution are distinct features. For instance, fractional Brownian motion exhibits Gaussian increments, while its MSD scales as a power law, whereas many diffusing diffusivity processes (72–75) exhibit non-Gaussian increments with a linear scaling of the MSD. Diffusion in such a passive crowded heterogeneous medium was formerly observed in non-living systems and called "anomalous (non-Gaussian) yet Brownian motion" (54,55) and the transition from subdiffusive regimes to normal diffusion with increasing lag time is commonly known (79,80). In contrast, the sweeping motion of microtubules creates local flows in the cytoplasm and results in a superdiffusive scaling in WT and noAct cases, $\langle r^2 \rangle_t \propto t^\alpha$, with an exponent $\alpha \approx 1.4$. While the superdiffusive motion of tracers in living cells was often observed (25,80,81,79,82,9), it was attributed to active bio-motors-driven transport along microtubules. In our experiment the mechanism is different. Moreover, we reveal for the first time both superdiffusive and non-Gaussian features of the passive intracellular transport in eukaryotic cells. Most importantly, we identify and decouple the origins of these fundamental aspects of intracellular dynamics.

**CONCLUSION**

In conclusion, the analysis of increment probability densities and autocorrelation functions reveals (a) the important role of the actin network reducing cell-to-cell heterogeneities and thus unifying intracellular material transport, (b) spatio-temporal heterogeneities as the reason for non-Gaussian transport features in living cells and (c) a scaling of the increment probability densities independent of the cytoskeleton state for 150 nm-diameter nanoparticles, mimicking the passive vesicle motion, up to the accessible measurement time. In this respect, the presence of the exponential tail for all four cytoskeleton states and the scaling of all increment probability density functions indicate a ubiquitous dynamic feature of intracellular material transport in the cytoplasm, independent of the cytoskeleton components present in the cell. These, in contrast, are responsible for regulating the efficiency of intracellular motion. Additionally, the investigated anomalous transport features reflect the heterogeneities within the cytoplasm and their spatio-temporal changes. These results pave the way for experimental investigation and theoretical analysis of intracellular material transport features based on spatio-temporal medium fluctuations, enabling future insights in cellular mechanisms.

**AUTHOR CONTRIBUTIONS**

D.H. and T.F. designed research; M.G. performed the experiments; P.W., Y.L., and D.G. analyzed data; and P.W., M.G., D.G., and D.H. wrote the paper. P.W. and M.G. contributed equally to the study. All authors discussed the results and commented on the manuscript.

**ACKNOWLEDGEMENTS**

P.W., M.G., and D.H. acknowledge Prof. Dr. Günther Gerisch (Max-Planck Institute for Biochemistry, Germany) for providing the *Dictyostelium discoideum* strains; P.W., M.G., and D.H. acknowledge funding from the Deutsche Forschungsgemeinschaft (grant HE5958-2-1), the Volkswagen-Foundation (grant I85100) and the Fraunhofer Attract Program for the grant "3DNanoCell"; DG and YL acknowledge the support of the French National Research Agency (Grant No. ANR-13-JSV5-0006-01).